\documentclass[aps,prl,showpacs,twocolumn,superscriptaddress, floatfix, nofootinbib]{revtex4-2}%

\usepackage[utf8]{inputenc}
\usepackage{epsfig}
\usepackage{url}
\usepackage{multirow}
\usepackage{makecell,booktabs} 
\usepackage{capt-of} 
\usepackage{graphicx}
\graphicspath{ {images/} }
\usepackage{tikz} 
\usepackage{amsfonts}
\usepackage{amssymb}
\usepackage{amsmath}
\usepackage{verbatim}
\usepackage{txfonts} 
\usepackage{float} 
\usepackage{bm} 
\usepackage{mathtools}
\usepackage{color,xcolor}
\newcommand*{\red}{\textcolor{black}} 
\usepackage{hyperref}
\hypersetup{  colorlinks=true, linkcolor=blue, citecolor=red, urlcolor=blue  }
\usepackage{physics}
\renewcommand{\bf}[1]{\mathbf{#1}}

\renewcommand{\eqref}[1]{Eq.~(\ref{#1})}

\begin{document}

\title{Experimental Simulation of Larger Quantum Circuits with Fewer Superconducting Qubits}

\author{Chong Ying}
\thanks{These authors contributed equally}
\affiliation{Hefei National Research Center for Physical Sciences at the Microscale and School of Physical Sciences, University of Science and Technology of China, Hefei 230026, China}
\affiliation{Shanghai Research Center for Quantum Science and CAS Center for Excellence in Quantum Information and Quantum Physics, University of Science and Technology of China, Shanghai 201315, China}
\affiliation{Hefei National Laboratory, University of Science and Technology of China, Hefei 230088, China}

\author{Bin Cheng}
\thanks{These authors contributed equally}
\affiliation{Centre for Quantum Software and Information, Faculty of Engineering and Information Technology, University of Technology Sydney, NSW 2007, Australia.}
\affiliation{Department of Physics, Southern University of Science and Technology, Shenzhen 518055, China.}
\affiliation{Shenzhen Institute for Quantum Science and Engineering, Southern University of Science and Technology, Shenzhen 518055, China.}

\author{Youwei Zhao}
\thanks{These authors contributed equally}
\affiliation{Hefei National Research Center for Physical Sciences at the Microscale and School of Physical Sciences, University of Science and Technology of China, Hefei 230026, China}
\affiliation{Shanghai Research Center for Quantum Science and CAS Center for Excellence in Quantum Information and Quantum Physics, University of Science and Technology of China, Shanghai 201315, China}
\affiliation{Hefei National Laboratory, University of Science and Technology of China, Hefei 230088, China}

\author{He-Liang Huang}
\email{quanhhl@ustc.edu.cn}
\affiliation{Hefei National Research Center for Physical Sciences at the Microscale and School of Physical Sciences, University of Science and Technology of China, Hefei 230026, China}
\affiliation{Shanghai Research Center for Quantum Science and CAS Center for Excellence in Quantum Information and Quantum Physics, University of Science and Technology of China, Shanghai 201315, China}
\affiliation{Hefei National Laboratory, University of Science and Technology of China, Hefei 230088, China}
\affiliation{Henan Key Laboratory of Quantum Information and Cryptography, Zhengzhou, Henan 450000, China}

\author{Yu-Ning Zhang}
\affiliation{Department of Physics, Southern University of Science and Technology, Shenzhen 518055, China.}
\affiliation{Shenzhen Institute for Quantum Science and Engineering, Southern University of Science and Technology, Shenzhen 518055, China.}
\affiliation{QuTech, Delft University of Technology, P.O. Box 5046, 2600 GA Delft, The Netherlands.}

\author{Ming Gong}
\email{minggong@ustc.edu.cn}
\author{Yulin Wu}
\affiliation{Hefei National Research Center for Physical Sciences at the Microscale and School of Physical Sciences, University of Science and Technology of China, Hefei 230026, China}
\affiliation{Shanghai Research Center for Quantum Science and CAS Center for Excellence in Quantum Information and Quantum Physics, University of Science and Technology of China, Shanghai 201315, China}
\affiliation{Hefei National Laboratory, University of Science and Technology of China, Hefei 230088, China}
\author{Shiyu Wang}
\author{Futian Liang}
\author{Jin Lin}
\author{Yu Xu}
\affiliation{Hefei National Research Center for Physical Sciences at the Microscale and School of Physical Sciences, University of Science and Technology of China, Hefei 230026, China}
\affiliation{Shanghai Research Center for Quantum Science and CAS Center for Excellence in Quantum Information and Quantum Physics, University of Science and Technology of China, Shanghai 201315, China}
\affiliation{Hefei National Laboratory, University of Science and Technology of China, Hefei 230088, China}
\author{Hui Deng}
\author{Hao Rong}
\author{Cheng-Zhi Peng}
\affiliation{Hefei National Research Center for Physical Sciences at the Microscale and School of Physical Sciences, University of Science and Technology of China, Hefei 230026, China}
\affiliation{Shanghai Research Center for Quantum Science and CAS Center for Excellence in Quantum Information and Quantum Physics, University of Science and Technology of China, Shanghai 201315, China}
\affiliation{Hefei National Laboratory, University of Science and Technology of China, Hefei 230088, China}

\author{Man-Hong Yung}
\email{yung@sustech.edu.cn}
\affiliation{Department of Physics, Southern University of Science and Technology, Shenzhen 518055, China.}
\affiliation{Shenzhen Institute for Quantum Science and Engineering, Southern University of Science and Technology, Shenzhen 518055, China.}
\affiliation{Guangdong Provincial Key Laboratory of Quantum Science and Engineering, Southern University of Science and Technology, Shenzhen 518055, China.}
\affiliation{Shenzhen Key Laboratory of Quantum Science and Engineering, Southern University of Science and Technology, Shenzhen, 518055, China.}

\author{Xiaobo Zhu}
\author{Jian-Wei Pan}
\affiliation{Hefei National Research Center for Physical Sciences at the Microscale and School of Physical Sciences, University of Science and Technology of China, Hefei 230026, China}
\affiliation{Shanghai Research Center for Quantum Science and CAS Center for Excellence in Quantum Information and Quantum Physics, University of Science and Technology of China, Shanghai 201315, China}
\affiliation{Hefei National Laboratory, University of Science and Technology of China, Hefei 230088, China}

\begin{abstract}
Although near-term quantum computing devices are still limited by the quantity and quality of qubits in the so-called NISQ era, quantum computational advantage has been experimentally demonstrated. Moreover, hybrid architectures of quantum and classical computing have become the main paradigm for exhibiting NISQ applications, where low-depth quantum circuits are repeatedly applied. In order to further scale up the problem size solvable by the NISQ devices, it is also possible to reduce the number of physical qubits by ``cutting" the quantum circuit into different pieces. 
In this work, we experimentally demonstrated a circuit-cutting method for simulating quantum circuits involving many logical qubits, using only a few physical superconducting qubits. By exploiting the symmetry of linear-cluster states, we can estimate the effectiveness of circuit-cutting for simulating up to 33-qubit linear-cluster states, using at most 4 physical qubits for each subcircuit. Specifically, for the 12-qubit linear-cluster state, we found that the experimental fidelity bound can reach as much as 0.734, which is about 19\% higher than a direct implementation {on the same} 12-qubit superconducting processor. Our results indicate that circuit-cutting represents a feasible approach of simulating quantum circuits using much fewer qubits, while achieving a much higher circuit fidelity. 
\end{abstract}

\maketitle

\textbf{Introduction.---}
Quantum computing offers potential speedups over classical computing on many applications, such as factoring~\cite{Shor-factorization,Lu2007Demonstration,huang2017Experimental}, unstructured search~\cite{Grover1996} and quantum simulation~\cite{Feynman1982-simulation,Lloyd1996-simulation,huang2021Emulating}. However, these applications require quantum computers to be fault-tolerant, which is still out of reach of the current quantum technology. Instead, we have just entered the \textit{noisy intermediate-scale quantum} (NISQ) era~\cite{Preskill2018,huang2020superconducting,huang2022near}, meaning that the number of physical qubits are sizable in terms of the computational space, but they are error-prone or noisy. Recent experimental demonstrations of quantum computation involves about 50 to 60 qubits~\cite{Arute2019,wu2021strong,zhu2022quantum,gong2022quantumneuronal}.
Although in terms of the memory size, they might have already exceeded the limits of classical super-computers, the incorrectable noisy gates limits the depth of quantum circuits running on current quantum devices, which constitutes a major obstacle to finding practical applications.
Therefore, it is of practical interest to solve large problems with smaller quantum devices, even with a tradeoff of using more classical resources.

This topic can be roughly categorized into two branches; one is at the algorithmic level, and the other is at the circuit level. 
The former is to decompose a large problem into smaller subproblems, each of which is solved by a small quantum computer. 
Examples include quantizing classical divide-and-conquer algorithms to solve combinatorial optimization problems~\cite{dunjko_computational_2018,ge_hybrid_2019}, and Fujii $et~al$'s deep variational quantum eigensolver framework~\cite{fujii_deep_2020}, which is suitable for simulating physical systems when interactions between subsystems are weak. 
Partially quantizing a tensor network may also fall into this category~\cite{liu_variational_2019, yuan_quantum_2020}.

\begin{figure}[t]
    \centering
    \includegraphics[width=0.37\textwidth]{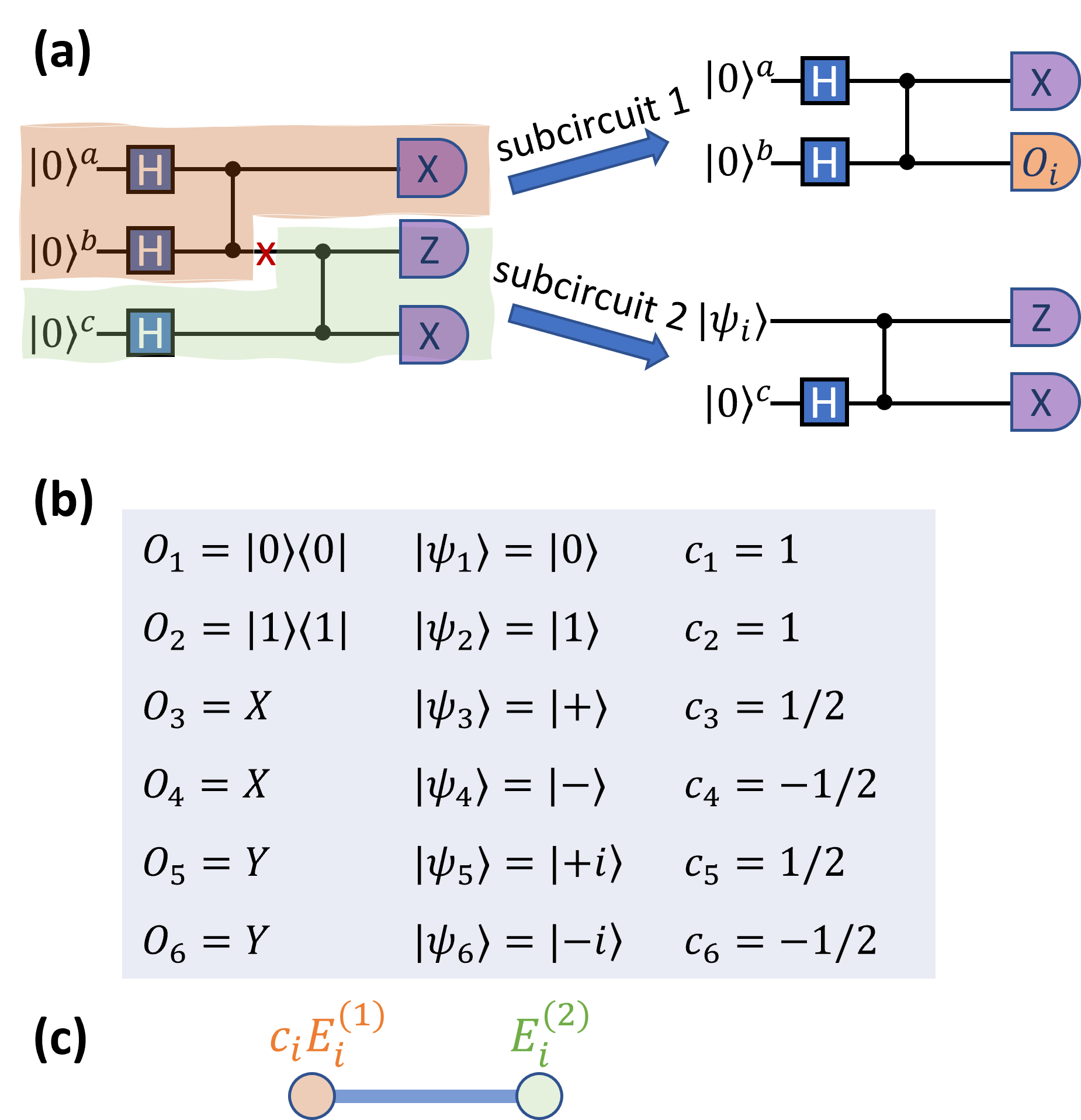}
    \caption{\textbf{(a)} Example illustrating the circuit-cutting scheme. The quantum circuit on the left is cut at the red cross, and partitioned into two subcircuits. \textbf{(b)} The list of $c_i, O_i$ and $\rho_i$ for \eqref{eq:rho_decomp}. \textbf{(c)} The tensor network representing the summation of \eqref{eq:example_expression}.}
    \label{fig:cutting}
\end{figure}

The circuit-level schemes intends to decompose a large quantum circuit into smaller pieces, implement each piece independently and finally use classical computers to combine \red{the} results.
For example, Bravyi $et~al.$~\cite{bravyi_trading_2016} discussed methods of using classical postprocessing to add virtual qubits for sparse circuits and Pauli-based computation. 
Mitarai and Fujii~\cite{mitarai_constructing_2019} proposed a method to add virtual two-qubit gates, which means that a remote two-qubit gate can be simulated by a quasiprobability decomposition of local single-qubit gates, thus cutting the large quantum circuit. 
Their work is for general quantum circuits and has been extended in a recent work~\cite{mitarai_overhead_2020} to allow decomposing non-local quantum channels into local ones.
On the other hand, using the language of tensor network, Peng $et$ $al.$~\cite{peng_simulating_2019} proposed a tomography-like circuit-cutting scheme, which is endowed with a rigorous analysis of the required quantum and classical resources to simulate general quantum circuits.
The circuit-cutting scheme is further analyzed and improved in later works~\cite{perlin_quantum_2020, ayral_quantum_2020, ayral_quantum_2021, tang_cutqc_2021}. 

In this work, we experimentally implement a tomography-like circuit-cutting scheme to simulate large linear-cluster states. 
The purpose of the experiment is to demonstrate the applicability and the actual performance of the circuit-cutting scheme in a real experimental platform.
Due to the symmetry in linear-cluster states, we only need to run subcircuits with at most 4 superconducting qubits and the simulated linear-cluster states scale up to 33 qubits. 
To analyze the performance, we use the stabilizer technique~\cite{toth2005entanglement,guhne_entanglement_2009} to estimate the fidelity lower bound. Then, it is compared with the fidelity bound obtained in a previous work that prepared the 12-qubit state directly~\cite{gong_genuine_2019}. 
The circuit-cutting scheme achieves a fidelity bound 0.734 in the 12-qubit case, which is about 19\% higher than the previous experiment.
Our experimental result shows the promise that the circuit-cutting scheme might become a standard tool in NISQ applications.

\textbf{Cutting large quantum circuits.---}
The basic idea is to cut a qubit wire and then simulate the propagation of quantum information by classical means. 
We illustrate this with a toy example in Fig.~\ref{fig:cutting}~(a). 
First, observe that at the time slice of the cutting point (the red cross), the reduced density matrix of the first two qubits can be decomposed as,
\begin{align}
    \rho^{ab} = \frac{1}{2} \sum_{j=0}^3 \Tr_b(\rho^{ab} \sigma_j^b) \otimes \sigma_j^b \ .
\end{align}
where we use superscripts to indicate the qubit labels and $\sigma_j \in \{ I, X, Y, Z \}$. Each Pauli operator can be further decomposed into its eigenstates, e.g., $Z = \dyad{0} - \dyad{1}$. 
Note that the identity operator can be written as $I = \dyad{0}+\dyad{1}$, which can be combined with $Z$~\cite{perlin_quantum_2020}. 
Then, we have,
\begin{align}\label{eq:rho_decomp}
    \rho^{ab} = \sum_{i=1}^6 c_i \Tr_b(\rho^{ab} O^b_i) \otimes \rho^b_i \ ,
\end{align}
where the $c_i, O_i$, and $\rho_i := \dyad{\psi_i}$ are listed in Fig.~\ref{fig:cutting}~(b). The partial trace operation in \eqref{eq:rho_decomp} can be interpreted as measuring $O_i$ in the qubit $b$ of subcircuit~1, and then $\rho_i$ is prepared and passed as input to subcircuit~2.

Suppose that we are interested in measuring the expectation of $X\otimes Z \otimes X$ of the 3-qubit circuit, denoted as $\ev{XZX}$.
In subcircuit~1, one needs to collect the expectation values of $X\otimes O_i$, defined by $E_i^{(1)} = \Tr( \rho^{ab} X\otimes O_i)$.
In subcircuit~2, one needs to collect the expectation values of $Z\otimes X$, denoted as $E_i^{(2)}$, from circuits with varying initial state $\ket{\psi_i}$ in the first qubit (see Fig.~\ref{fig:cutting}~(a)). Then, according to \eqref{eq:rho_decomp}, $\ev{XZX}$ can be recovered by~\cite{peng_simulating_2019}
\begin{align}\label{eq:example_expression}
    \ev{XZX} = \sum_{i=1}^6 c_i E_i^{(1)} E_i^{(2)} \ .
\end{align}
This circuit-cutting procedure works for any observable in the form $A\otimes B$, where $A$ is an observable of the qubit~$a$ and $B$ is an observable of the qubits~$b$ and $c$.
We remark that the combination of expectation values is achieved with a classical computer. 
In this process, we do not create a 3-qubit entangled state; instead, the 3-qubit state is simulated by a hybrid scheme of a 2-qubit quantum computer and a classical computer.

For more general and larger quantum circuits, one can apply this cutting scheme iteratively to multiple cutting points, to partition the whole circuit into several \emph{disconnected} pieces of subcircuits. 
By running the subcircuits \emph{independently}, and classically combining the subcircuit expectations with appropriate coefficients, one obtains expectations from the large quantum circuits. Moreover, those disconnected subcircuits can be viewed as nodes in a tensor network~\cite{peng_simulating_2019}.
For example, the corresponding tensor network for Fig.~\ref{fig:cutting}~(a) is a line with two nodes, and the edge has bond dimension 6, corresponding to the 6 terms in \eqref{eq:example_expression}.
The coefficient $c_i$ can be absorbed into the node representing $E_i^{(1)}$ or $E_i^{(2)}$. Then, one can use tensor-network contraction to perform the combination to obtain quantities of the large circuit, with classical running time exponential in the treewidth of the tensor network~\cite{peng_simulating_2019}.

To summarize, the protocol is as follows. 
\emph{(a)} Identify appropriate cutting points to partition the large circuit into disconnected subcircuits. 
\emph{(b)} Obtain the subcircuit expectations by enumerating the possible choices of $\ket{\psi_i}$ and $O_i$. 
\emph{(c)} Construct a tensor network from these subcircuit expectations and coefficients $c_i$.
\emph{(d)} Contract the tensor network to obtain the expectation value with respect to the large circuit.

\textbf{Linear-cluster states.---}
Cluster states are a family of highly-entangled states, which can be used to achieve measurement-based quantum computation~\cite{Raussendorf2001one,nielsen2006cluster}. Linear-cluster state is a specific example of cluster states, where all qubits are aligned in one dimension. 
Explicitly, a linear-cluster state with $n$ qubits can be expressed as,
\begin{align}\label{eq:LC_state_def}
    \ket{\mathsf{LC}_n} = \left( \prod_{i=1}^{n-1} \mathrm{CZ}^{i, i+1} \right) \ket{+}^{\otimes n} \ ,
\end{align}
where the superscripts in the CZ gates indicate the qubits that they act on.

\begin{figure}
    \centering
    \includegraphics[width=0.43\textwidth]{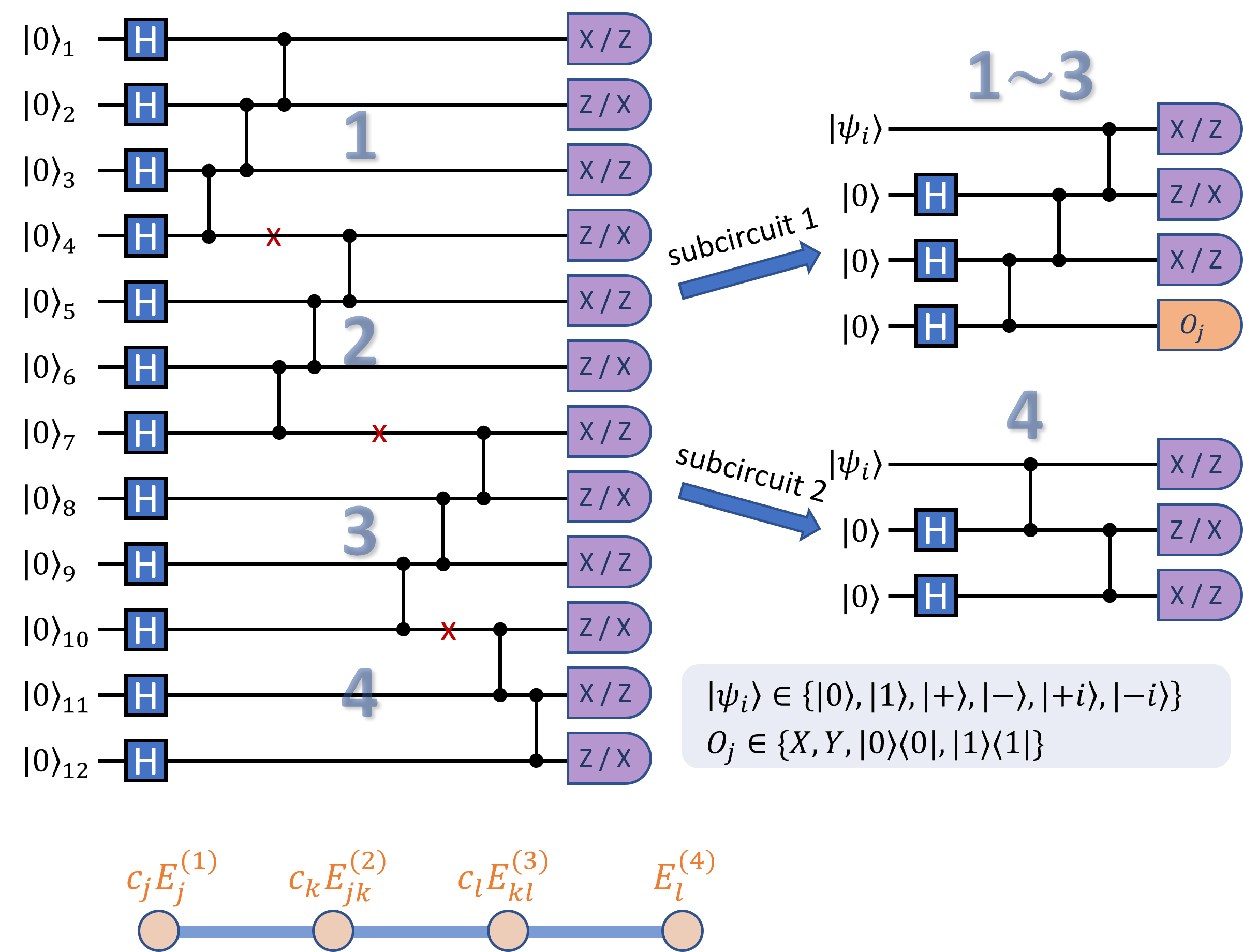}
    \caption{Circuit-cutting scheme for the linear-cluster state. 
    \textit{Left.} A 12-qubit linear-cluster state, which is cut into 4 pieces. 
    \textit{Right.} The 12-qubit linear-cluster state can be simulated by combining measurement data from these two \red{types of} subcircuits.
    \textit{Bottom.} The tensor network representing the classical combination of subcircuits.}
    \label{fig:exp-picture}
\end{figure}

In this work, we experimentally simulate a 12-qubit linear-cluster state, with 4 qubits of a superconducting quantum processor, which is the same processor as in Ref.~\cite{gong_genuine_2019}. 
As in Fig.~\ref{fig:exp-picture}, there are 3 cutting points on the 12-qubit circuit, partitioning it into 4 subcircuits. The first 3 subcircuits are all in the form of subcircuit~1, while the last subcircuit is in the form of subcircuit~2. Note that the sequence of CZ gates on the left of Fig.~\ref{fig:exp-picture} is chosen such that pieces 1-3 can be represented by the same subcircuit~1.

To compare the performance of the circuit cutting scheme with that of running the 12-qubit circuit directly, we need to estimate their fidelities. 
We follow the approach in Ref.~\cite{gong_genuine_2019}, which uses techniques from entanglement detection in the stabilizer formalism~\cite{toth2005entanglement,guhne_entanglement_2009}.
Let $s_1 = X_1 Z_2$, $s_n = Z_{n-1} X_n$ and $s_i = Z_{i-1} X_i Z_{i+1}$ for $i \neq 1$ or $n$.
It can be shown that a linear-cluster state is a stabilizer state with a stabilizer group spanned by $\{s_1, \cdots, s_n \}$, i.e., $s_i \ket{\mathsf{LC}_n} = \ket{\mathsf{LC}_n}$ for $i=1, \cdots, n$.
Let
\begin{align}\label{eq:odd_and_even}
    \mathsf{ODD}_n &:= \prod_{i \text{ odd}} \frac{1 + s_i}{2} & \mathsf{EVEN}_n &:= \prod_{i \text{ even}} \frac{1 + s_i}{2} \ .
\end{align}
For a linear-cluster state, one has $\dyad{\mathsf{LC}_n} \geq \mathsf{ODD}_n + \mathsf{EVEN}_n - I$~\cite[Theorem 6]{toth2005entanglement}.
Therefore, for an unknown quantum state $\rho$, its fidelity relative to the linear-cluster state is lower bounded by,
\begin{align}\label{eq:fidelity_lower_bound}
    \Tr(\rho \dyad{\mathsf{LC}_n}) \geq \Tr(\rho \mathsf{ODD}_n) + \Tr(\rho \mathsf{EVEN}_n) - 1 \ ,
\end{align}
which can be estimated by measuring $\mathsf{ODD}_n$ and $\mathsf{EVEN}_n$.
Observe that every term in the expansion of $\mathsf{ODD}_n$ can be measured in the basis $XZXZ\cdots$, while every term in the expansion of $\mathsf{EVEN}_n$ can be measured in the basis $ZXZX\cdots$. 
Therefore, to estimate the fidelity, one only needs to perform measurements in two bases.
For simplicity, we will refer to them as $XZ$ measurement and $ZX$ measurement, respectively.

Below, we illustrate how to simulate the 12-qubit linear-cluster state with the circuit-cutting scheme.
Suppose we want to obtain the expectation value $\ev*{P^{(1)} \otimes P^{(2)} \otimes P^{(3)} \otimes P^{(4)}}$, where $P^{(i)}$ can be any \red{3-qubit} observable \red{of the $i$-th 3-qubit group}. In Fig.~\ref{fig:exp-picture}, we denote the final state of subcircuit 1 and 2 as $\ket{\Phi_{1, i}}$ and $\ket{\Phi_{2, i}}$, respectively, where the index $i$ indicates one of the 6 states $\ket{\psi_i}$ in the first qubit.
Then, define
\begin{align}
    E^{(1)}_{j} &:= \mel*{\Phi_{1, 3}}{P^{(1)} \otimes O_j}{\Phi_{1, 3}} & E^{(2)}_{jk} &:= \mel*{\Phi_{1, j}}{P^{(2)} \otimes O_k}{\Phi_{1, j}} \notag \\
    E^{(3)}_{kl} &:= \mel*{\Phi_{1, k}}{P^{(3)} \otimes O_l}{\Phi_{1, k}} & E^{(4)}_l &:= \mel*{\Phi_{2, l}}{P^{(4)}}{\Phi_{2, l}}
\end{align}
to be the subcircuit expectations from pieces 1-4.
According to the circuit-cutting scheme, we have,
\begin{align}\label{eq:12_qubit_circuit_cutting}
    \ev*{P^{(1)} \otimes P^{(2)} \otimes P^{(3)} \otimes P^{(4)}} = \sum_{j, k, l = 1}^6 c_j c_k c_l E^{(1)}_{j} E^{(2)}_{jk} E^{(3)}_{kl} E^{(4)}_l  \ ,
\end{align}
\red{where the coefficients $c_i$'s are shown in Fig.~\ref{fig:cutting}~(b).}
Again, this summation can be viewed as tensor network contraction as in the bottom of Fig.~\ref{fig:exp-picture}.

The experimental procedure \red{for estimating the fidelity} is as follows. 
(\emph{a}) Identify the observables in the expansion of $\mathsf{ODD}_{12}$ and $\mathsf{EVEN}_{12}$.
(\emph{b}) For each observable, define $P^{(i)}$ for $i = 1,2,3,4$.
Measure $E^{(1)}_{j}$, $E^{(2)}_{jk}$, $E^{(3)}_{kl}$ and $E^{(4)}_l$, and use \eqref{eq:12_qubit_circuit_cutting} to obtain the expectation value of that observable.
(\emph{c}) Calculate the fidelity lower bound according to \eqref{eq:fidelity_lower_bound}.
Note that this procedure can be easily generalize to larger linear-cluster states.

\red{
The expectation values in \eqref{eq:12_qubit_circuit_cutting} can be obtained from subcircuits in Fig.~\ref{fig:exp-picture}.
For subcircuit~1, we need to prepare the circuits with 6 different $\ket{\psi_i}$.
The measurement bases for the first three qubits are $XZX$ and $ZXZ$, and for the last qubit are $X, Y$ or $Z$; the expectation value of $\dyad{0}$ or $\dyad{1}$ can be obtained from $Z$ measurement.
Therefore, we need to implement $6\times 2\times 3 = 36$ different circuits in the form of subcircuit~1.
As for subcircuit~2, similar argument shows that we need to implement $12$ different circuits.
Thus, a total of 48 subcircuits needs to be run~\cite{supp}.}

\begin{figure}[t]
    \centering
    \includegraphics[width=0.43\textwidth]{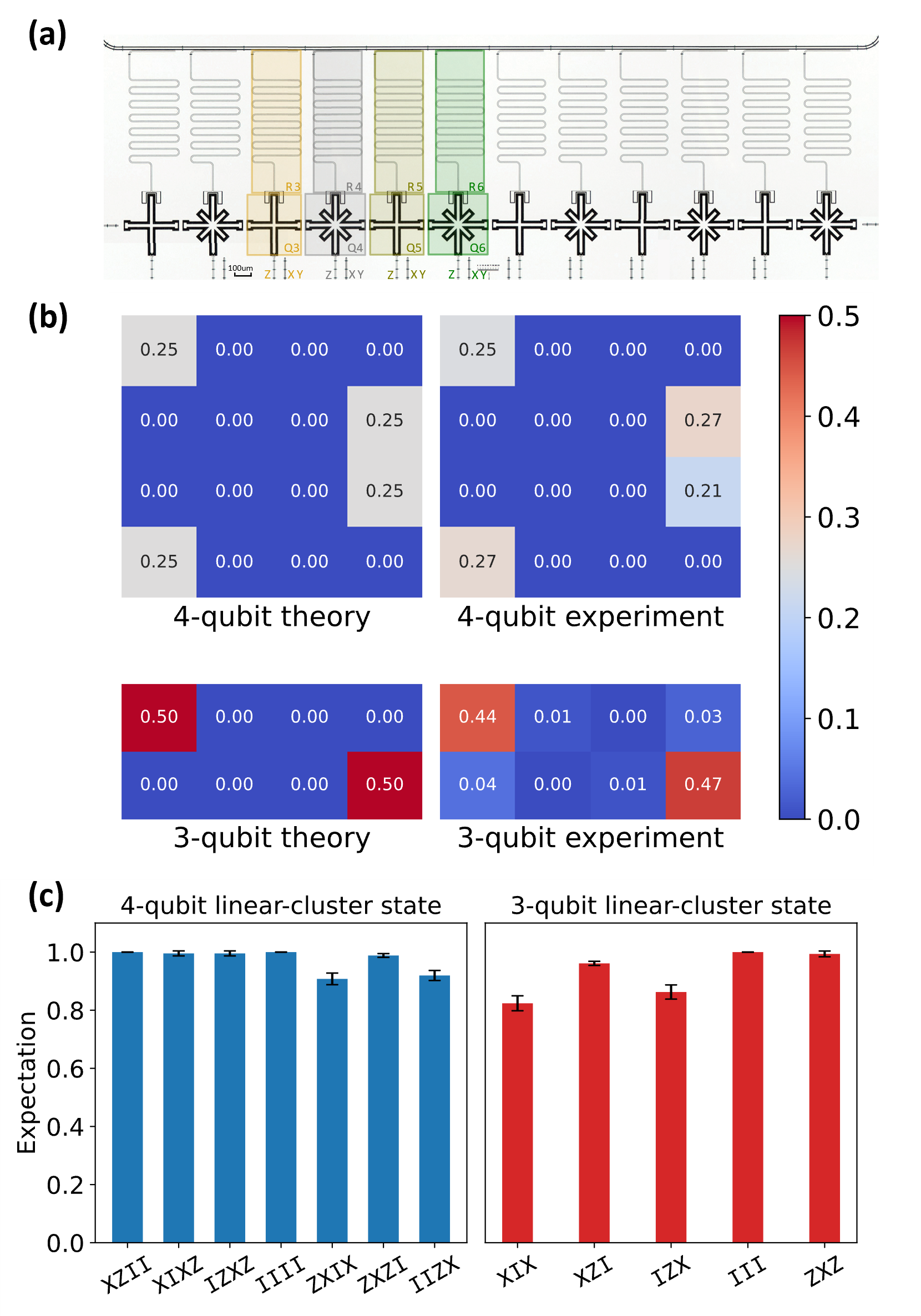}
    \caption{\textbf{(a)} Schematic of the 12-qubit superconducting processor, where we used Q3 to Q6 for the circuit-cutting experiment.
    \textbf{(b)} The output distributions from $XZ$ measurement of the 4-qubit and 3-qubit linear-cluster states; \red{see Supplemental Material~\cite{supp} for the labelling of each cell.}
    \textbf{(c)} Expectation from $XZ$ and $ZX$ measurements of the 4-qubit and 3-qubit linear-cluster states. Ideal values are one. The error bars are due to the repeated experiments.}
    \label{fig:subcirc_data}
\end{figure}

\textbf{Experiment.---}
To verify the feasibility and evaluate the actual performance of the scheme in the experiment, we run the subcircuits in Fig.~\ref{fig:exp-picture} on a 12-qubit superconducting quantum processor. 
As shown in Fig.~\ref{fig:subcirc_data}~(a), the qubits are arranged in a one-dimensional chain. Each qubit has two control lines
to provide full control of the qubit: a microwave $XY$ control line
to drive excitations between $\vert0\rangle$ and $\vert1\rangle$, and a magnetic flux
bias line to tune the qubit resonance frequency. As the near-neighbor qubits are capacitively coupled, the fast adiabatic CZ gates~\cite{barends2014superconducting,martinis2014fast} can be applied. The measurements of qubit are done through dispersively coupling to a readout resonator. We choose four adjacent qubits from a 12-qubit superconducting quantum processor to implement
the experiments. The average performance of the chosen qubits are: $T_1 \approx$ 36.1 $\mu$s, $T_2^* \approx$ 4.3 $\mu$s, single-qubit gate fidelity $\approx$ 99.93\% and CZ gate fidelity $\approx$ 98.5\%. More detailed data are shown in the Supplemental Material~\cite{supp}.

All the experimental results are processed using the transition matrix error mitigation (TMEM) method~\cite{PhysRevA.103.042605,PhysRevLett.127.090502}, to suppress the readout noise. 
However, negative entries may appear in the probability distributions of the subcircuits after the TMEM. 
To make these distributions physical, we first transform those probability distributions into diagonal operators, and then use the maximum likelihood method to find a density operator that is the closest to them~\cite{maciejewski2020mitigation, james2005measurement}.
The final distributions of the subcircuits are then extracted from these density operators.
Before and after the experiment of circuit cutting, additional quantum state tomography on the final state of the circuit is performed to evaluate the performance of the experiments. 
The average fidelity of the 36 \red{subcircuits in the form of subcircuit~1} is 0.944, and the average fidelity of the 12 \red{subcircuits in the form of subcircuit~2} is 0.955, showing the high quality of the experiments.

\begin{figure}[t]
    \centering
    \includegraphics[width=0.45\textwidth]{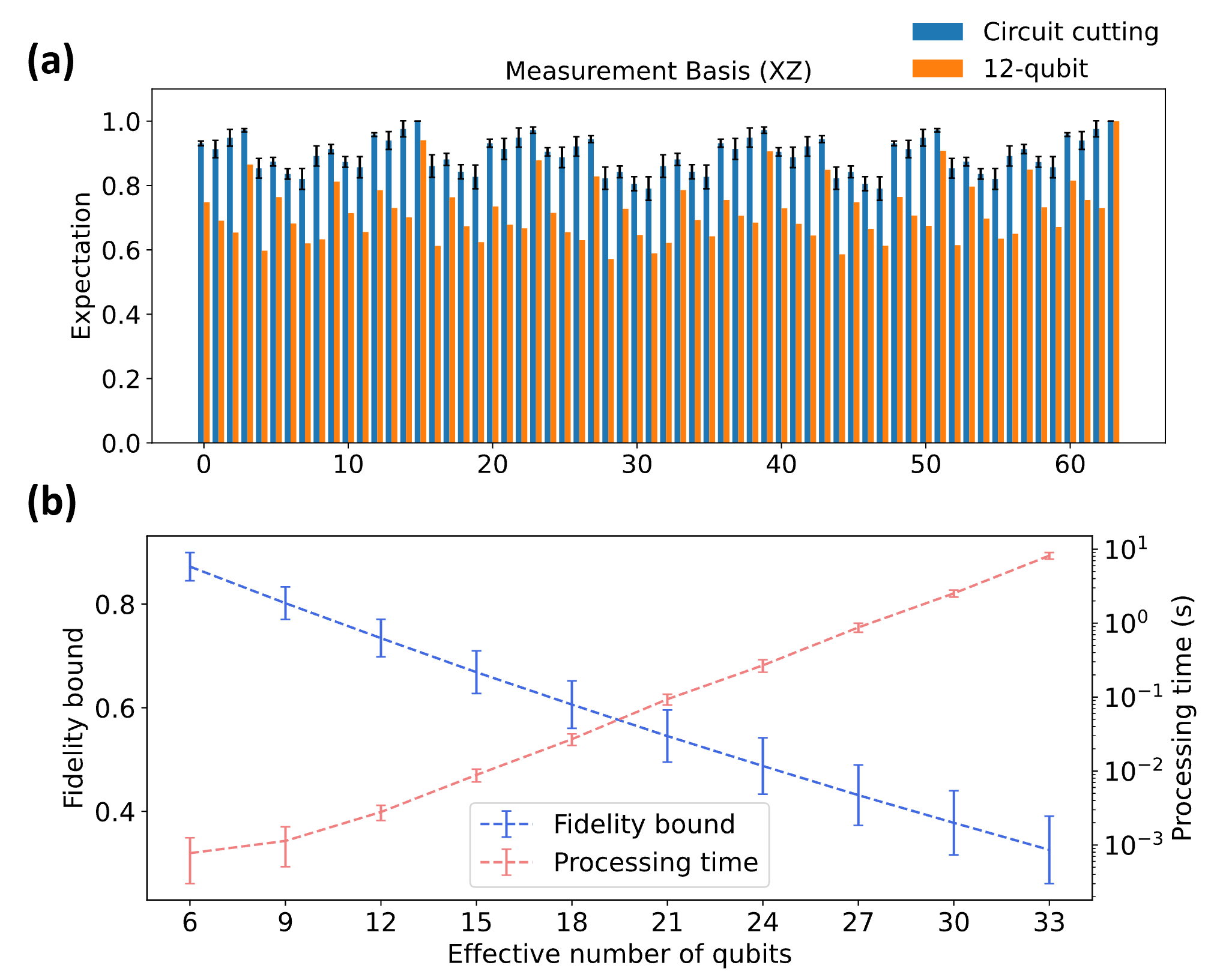}
    \caption{\textbf{(a)} Comparison of the expectations with $XZ$ measurement obtained from the 12-qubit circuit~\cite{gong_genuine_2019} (orange) and from the circuit-cutting scheme (blue). \textbf{(b)} Fidelity bound (blue) and processing time (red) for simulating larger linear-cluster state using the same experimental data. 
    The error bars are due to repeated experiments.}
    \label{fig:data_exp}
\end{figure}

As a warm-up, we show how to estimate the fidelity lower bounds for the 4-qubit and 3-qubit linear-cluster states. Note that if we take $\ket{\psi_i} = \ket{+}$ for the subcircuits, then they correspond to a 4-qubit and 3-qubit linear-cluster state, respectively.
The corresponding probability distributions from the $XZ$ measurement are shown in Fig.~\ref{fig:subcirc_data}~(b), which also includes the theoretical distributions for a comparison;
\red{see Supplemental Material~\cite{supp} for the probability distributions from $ZX$ measurement.} \red{From these distributions, one can obtain the expectations of terms in the $\mathsf{ODD}$ and $\mathsf{EVEN}$ operators of the 4-qubit and 3-qubit LC states, as shown in Fig.~\ref{fig:subcirc_data}~(c).
The fidelity lower bound then follows from these expectations according to \eqref{eq:fidelity_lower_bound}, which is 0.952 and 0.909 for the 4-qubit and 3-qubit LC states, respectively~\cite{supp}.}
These bounds match the average fidelity \red{(of all 36 circuits for subcircuit~1 and 12 circuits for subcircuit~2)} from quantum state tomography. 

We now turn our discussion to simulating large linear-cluster state with the circuit-cutting scheme.
With the measurement data from the subcircuits, one can simulate larger linear-cluster states, and the fidelity bounds can be derived with similar procedures. 
Fig.~\ref{fig:data_exp}~(a) presents the expectations of terms in $\mathsf{ODD}_{12}$ for the 12-qubit state obtained by the circuit-cutting scheme and a direct implementation, \red{which is an analogue of Fig.~\ref{fig:subcirc_data}~(c)}.
Those expectations are from the $XZ$ measurement and there are $2^6$ expectations in total.
The blue bars are reconstructed from the circuit-cutting scheme, while the orange bars are from the experimental data in Ref.~\cite{gong_genuine_2019}. 
Each bar corresponds to one specific terms in the expansion of $\mathsf{ODD}_{12}$ and the ideal value is one; \red{we leave the concrete labelling and similar data for $\mathsf{EVEN}_{12}$ in Supplemental Materials~\cite{supp}}. We remark that the distributions from the 12-qubit experiment are also processed with the same procedure (the fidelity bound after processing is 0.615), i.e., TMEM followed by a maximum likelihood method, for a fair comparison.
From these expectations, we can similarly use \eqref{eq:fidelity_lower_bound} to estimate the fidelity.
The estimated fidelity bound from the circuit-cutting scheme is 0.734, about 19\% higher than that from the experiment of Ref.~\cite{gong_genuine_2019}.
\red{Note that the experiment of Ref.~\cite{gong_genuine_2019} implemented CZ gates in parallel, which will incur more severe crosstalk errors compared to our current implementation, where CZ gates are applied individually (one for each layer; see the right of Fig.~\ref{fig:exp-picture}).
Moreover, smaller circuits are easier to calibrate and control.
Therefore, the circuit-cutting experiment achieves a better fidelity bound than Ref.~\cite{gong_genuine_2019}.}
The tradeoff for the circuit-cutting scheme is the increase of both the quantum and the classical running time~\cite{supp}.

Moreover, the symmetry in linear-cluster states allows us to reuse the measurement data from subcircuit~1 to simulate larger linear-cluster states, at a cost of increasing overhead in classical postprocessing. 
Specifically, we need to add more internal nodes to the tensor network in Fig.~\ref{fig:exp-picture}, to represent larger circuits (5 nodes for 15 qubits, 6 nodes for 18 qubits and so on).
This allows us to simulate linear-cluster states of size $6 + 3 k$, where $k$ is a positive integer.
We need to contract a longer chain to obtain one expectations of the large circuit, and there will be more expectations to be computed in order to obtain the fidelity lower bound.
The obtained fidelity bound is expected to decay as the number of qubits increases, since the error accumulates in the classical postprocessing. 
The fidelity decay and classical postprocessing time are shown in Fig.~\ref{fig:data_exp}~(b).
Here, the classical postprocessing is done on a conventional laptop, and the processing time shows the running time of the program for calculating the fidelity lower bound of larger circuits~\cite{data_and_code}.

\textbf{Discussion.---}
In this work, we experimentally demonstrate a circuit-cutting scheme and simulate larger linear-cluster state with size scaling up to 33 qubits, using at most 4 qubits. 
In the case of 12 qubits, we achieve a higher fidelity compared to that of a previous work that prepared the 12-qubit state directly~\cite{gong_genuine_2019}, giving supportive evidence to the applicability of the circuit-cutting scheme. 

Simulating large quantum circuits with small quantum devices is a promising direction in the NISQ era. Currently, there exist several circuit-cutting schemes~\cite{bravyi_trading_2016, peng_simulating_2019, mitarai_constructing_2019, mitarai_overhead_2020}; it is necessary to further perform experimental benchmarking on these schemes, in order to evaluate their applicability in practice. On the other hand, although circuit-cutting schemes provide systematic methods to cutting quantum circuits into smaller pieces, to the best of our knowledge, there is no general method for determining the optimal cutting points. Therefore, we believe that the potential of circuit-cutting has not yet been fully explored.

\begin{acknowledgments}
\textbf{Acknowledgement.---}
The authors thank the Laboratory of Microfabrication, University of Science and Technology of China, Institute of Physics CAS, and National Center for Nanoscience and Technology for supporting the sample fabrication. The authors also thank QuantumCTek Co., Ltd., for supporting the fabrication and the maintenance of room-temperature electronics.
BC thanks the support from the Sydney Quantum Academy, Sydney, NSW, Australia.
\textbf{Funding}: This research was supported by Innovation Program for Quantum Science and Technology (Grant No.~2021ZD0300200), NSFC (Grants No. 11574380, No. 11905217), the Chinese Academy of Science and its Strategic Priority Research Program (Grants No. XDB28000000), Shanghai Municipal Science and Technology Major Project (Grant No.2019SHZDZX01), the Science and Technology Committee of Shanghai Municipality, Anhui Initiative in Quantum Information Technologies, and Special funds from Jinan science and Technology Bureau and Jinan high tech Zone Management Committee. MHY is supported by the National Natural Science Foundation of China (11875160), the Guangdong Innovative and Entrepreneurial Research Team Program (2016ZT06D348), Natural Science Foundation of Guangdong Province (2017B030308003), and Science, Technology and Innovation Commission of Shenzhen Municipality (ZDSYS20170303165926217, JCYJ20170412152620376, JCYJ20170817105046702). H.-L. H. acknowledges support from the Youth Talent Lifting Project (Grant No. 2020-JCJQ-QT-030), National Natural Science Foundation of China (Grants No. 11905294, 12274464), China Postdoctoral Science Foundation, and the Open Research Fund from State Key Laboratory of High Performance Computing of China (Grant No. 201901-01).
\end{acknowledgments}

\bibliography{ref}

\end{document}


\title{Supplemental Material for ``Experimental Simulation of Larger Quantum Circuits with Fewer Superconducting Qubits''}

\author{Chong Ying}
\thanks{These authors contributed equally}
\affiliation{Hefei National Research Center for Physical Sciences at the Microscale and School of Physical Sciences, University of Science and Technology of China, Hefei 230026, China}
\affiliation{Shanghai Research Center for Quantum Science and CAS Center for Excellence in Quantum Information and Quantum Physics, University of Science and Technology of China, Shanghai 201315, China}
\affiliation{Hefei National Laboratory, University of Science and Technology of China, Hefei 230088, China}

\author{Bin Cheng}
\thanks{These authors contributed equally}
\affiliation{Centre for Quantum Software and Information, Faculty of Engineering and Information Technology, University of Technology Sydney, NSW 2007, Australia.}
\affiliation{Department of Physics, Southern University of Science and Technology, Shenzhen 518055, China.}
\affiliation{Shenzhen Institute for Quantum Science and Engineering, Southern University of Science and Technology, Shenzhen 518055, China.}

\author{Youwei Zhao}
\thanks{These authors contributed equally}
\affiliation{Hefei National Research Center for Physical Sciences at the Microscale and School of Physical Sciences, University of Science and Technology of China, Hefei 230026, China}
\affiliation{Shanghai Research Center for Quantum Science and CAS Center for Excellence in Quantum Information and Quantum Physics, University of Science and Technology of China, Shanghai 201315, China}
\affiliation{Hefei National Laboratory, University of Science and Technology of China, Hefei 230088, China}

\author{He-Liang Huang}
\email{quanhhl@ustc.edu.cn}
\affiliation{Hefei National Research Center for Physical Sciences at the Microscale and School of Physical Sciences, University of Science and Technology of China, Hefei 230026, China}
\affiliation{Shanghai Research Center for Quantum Science and CAS Center for Excellence in Quantum Information and Quantum Physics, University of Science and Technology of China, Shanghai 201315, China}
\affiliation{Hefei National Laboratory, University of Science and Technology of China, Hefei 230088, China}
\affiliation{Henan Key Laboratory of Quantum Information and Cryptography, Zhengzhou, Henan 450000, China}

\author{Yu-Ning Zhang}
\affiliation{Department of Physics, Southern University of Science and Technology, Shenzhen 518055, China.}
\affiliation{Shenzhen Institute for Quantum Science and Engineering, Southern University of Science and Technology, Shenzhen 518055, China.}
\affiliation{QuTech, Delft University of Technology, P.O. Box 5046, 2600 GA Delft, The Netherlands.}

\author{Ming Gong}
\email{minggong@ustc.edu.cn}
\author{Yulin Wu}
\affiliation{Hefei National Research Center for Physical Sciences at the Microscale and School of Physical Sciences, University of Science and Technology of China, Hefei 230026, China}
\affiliation{Shanghai Research Center for Quantum Science and CAS Center for Excellence in Quantum Information and Quantum Physics, University of Science and Technology of China, Shanghai 201315, China}
\affiliation{Hefei National Laboratory, University of Science and Technology of China, Hefei 230088, China}
\author{Shiyu Wang}
\author{Futian Liang}
\author{Jin Lin}
\author{Yu Xu}
\affiliation{Hefei National Research Center for Physical Sciences at the Microscale and School of Physical Sciences, University of Science and Technology of China, Hefei 230026, China}
\affiliation{Shanghai Research Center for Quantum Science and CAS Center for Excellence in Quantum Information and Quantum Physics, University of Science and Technology of China, Shanghai 201315, China}
\affiliation{Hefei National Laboratory, University of Science and Technology of China, Hefei 230088, China}
\author{Hui Deng}
\author{Hao Rong}
\author{Cheng-Zhi Peng}
\affiliation{Hefei National Research Center for Physical Sciences at the Microscale and School of Physical Sciences, University of Science and Technology of China, Hefei 230026, China}
\affiliation{Shanghai Research Center for Quantum Science and CAS Center for Excellence in Quantum Information and Quantum Physics, University of Science and Technology of China, Shanghai 201315, China}
\affiliation{Hefei National Laboratory, University of Science and Technology of China, Hefei 230088, China}

\author{Man-Hong Yung}
\email{yung@sustech.edu.cn}
\affiliation{Department of Physics, Southern University of Science and Technology, Shenzhen 518055, China.}
\affiliation{Shenzhen Institute for Quantum Science and Engineering, Southern University of Science and Technology, Shenzhen 518055, China.}
\affiliation{Guangdong Provincial Key Laboratory of Quantum Science and Engineering, Southern University of Science and Technology, Shenzhen 518055, China.}
\affiliation{Shenzhen Key Laboratory of Quantum Science and Engineering, Southern University of Science and Technology, Shenzhen, 518055, China.}

\author{Xiaobo Zhu}
\author{Jian-Wei Pan}
\affiliation{Hefei National Research Center for Physical Sciences at the Microscale and School of Physical Sciences, University of Science and Technology of China, Hefei 230026, China}
\affiliation{Shanghai Research Center for Quantum Science and CAS Center for Excellence in Quantum Information and Quantum Physics, University of Science and Technology of China, Shanghai 201315, China}
\affiliation{Hefei National Laboratory, University of Science and Technology of China, Hefei 230088, China}

\maketitle

\onecolumngrid

\beginsupplement

\section{Experimental Details}

Our superconducting quantum processor has 12 frequency-tunable transmon qubits lying on a 1D chain, and the qubits are capacitively coupled to their nearest neighbors (the coupling strength is about 12 MHz). Each qubit has a microwave drive line (XY), a fast flux-bias line (Z) and a readout resonator. All readout resonators are coupled to a common transmission line for state readout. In our experiments, four qubits are chosen from the quantum processor. The performances of the four qubits we chosen in our experiment are listed in Table \ref{tab:chip-performance}. Since we blocked the other qubits by tuning their frequencies far away, the performance of the selected 4 qubits is obviously improved after fine calibartion, compared with the same qubits performed in the 12-qubit experiment. It is much more difficult to calibrate 12 qubits to work well simultaneously, due to the problems such as frequency crowding and crosstalk, which are generally encountered in large scale superconducting quantum processors.

\begin{table*}[htbp]
\begin{tabular}{lcccc}
\toprule
  & $Q_3$ & $Q_4$ & $Q_5$ & $Q_6$ \\
\midrule
$\omega_{01}$ (GHz) & 4.916 & 4.352 & 5.081 & 4.195  \\
$\eta$ (MHz) & -246 & -203 & -247 & -202  \\
$T_1$ ($\mu$s) & 45.7 & 35.7 & 29.8 & 33.5  \\
$T_2^*$ ($\mu$s) & 3.99 & 2.93 & 8.14 & 2.33  \\
$f_{00}$ (\%) & 95.0  & 94.3  & 96.9  & 92.2  \\
$f_{11}$ (\%) & 90.9  & 91.0  & 90.1  & 88.7  \\
single-qubit gate fidelity (\%) & 99.93$\pm$0.07  & 99.97$\pm$0.10  & 99.90$\pm$0.08  & 99.91$\pm$0.17  \\
CZ fidelity (\%) & \multicolumn{4}{l}{~~~~~~~~~~~~97.3$\pm$1.7   ~~~~~~~98.9$\pm$2.0  ~~~~ 99.2$\pm$2.2} \\
\bottomrule
\end{tabular}
\caption{Performance of qubits. $\omega_{01}$ is the idle frequency of qubit. $\eta$ is the anharmonicity. $T_1$ is the energy relaxation time. $T_2^*$ is the dephasing time extracted from Ramsey experiment. $f_{00}$ ($f_{11}$) is the probability of correctly readout of qubit state in $\ket{0}$ ($\ket{1}$) after being well prepared in $\ket{0}$ ($\ket{1}$). Single-qubit gate fidelities and CZ gate fidelities are characterized by randomization benchmarking. }
\label{tab:chip-performance}
\end{table*}

\section{Circuit-cutting Process}
For a mixed system $abc$ where $b$ denotes the index of a single qubit, the density matrix can be decomposed as,
\begin{align}
  {{\rho }^{abc}}=& \text{T}{{\text{r}}_{\text{b}}}\left( {{\rho }^{ab}}\left| 0 \right\rangle \left\langle  0 \right| \right)\otimes \rho _{\left| 0 \right\rangle }^{bc}+\text{T}{{\text{r}}_{\text{b}}}\left( {{\rho }^{ab}}\left| 1 \right\rangle \left\langle  1 \right| \right)\otimes \rho _{\left| 1 \right\rangle }^{bc}+\frac{1}{2}\text{T}{{\text{r}}_{\text{b}}}\left( {{\rho }^{ab}}X \right)\otimes \rho _{\left| + \right\rangle }^{bc} \notag\\
  & -\frac{1}{2}\text{T}{{\text{r}}_{\text{b}}}\left( {{\rho }^{ab}}X \right)\otimes \rho _{\left| - \right\rangle }^{bc}+\frac{1}{2}\text{T}{{\text{r}}_{\text{b}}}\left( {{\rho }^{ab}}Y \right)\otimes \rho _{\left| +i \right\rangle }^{bc}-\frac{1}{2}\text{T}{{\text{r}}_{\text{b}}}\left( {{\rho }^{ab}}Y \right)\otimes \rho _{\left| -i \right\rangle }^{bc}  
\end{align}
where $\rho_{\psi}^{bc}$  denotes the density matrix of subsystem $bc$ with qubit $b$ initiated in state $\psi$.

In the experiment, the diagonal term of the density matrix is often concerned and accessible directly. Let's define ${{\rho }_{\alpha \beta \gamma }}\equiv {{\rho }_{\alpha \beta \gamma ,\alpha \beta \gamma }}$ to denote the $\alpha \beta \gamma th$ diagonal term of the density matrix for the sake of brevity. Therefor, the diagonal term of the mixed system can be expressed as,
\begin{align}
   \rho _{\alpha \beta \gamma }^{abc}=\rho _{\alpha 0}^{ab}\rho _{\left| 0 \right\rangle ,\beta \gamma }^{bc}+\rho _{\alpha 1}^{ab}\rho _{\left| 1 \right\rangle ,\beta \gamma }^{bc}+\frac{1}{2}(\rho _{\alpha 0}^{X,ab}-\rho _{\alpha 1}^{X,ab})(\rho _{\left| + \right\rangle ,\beta \gamma }^{bc}-\rho _{\left| - \right\rangle ,\beta \gamma }^{bc})  +\frac{1}{2}(\rho _{\alpha 0}^{Y,ab}-\rho _{\alpha 1}^{Y,ab})(\rho _{\left| +i \right\rangle ,\beta \gamma }^{bc}-\rho _{\left| -i \right\rangle ,\beta \gamma }^{bc})  
\end{align}
where $\alpha$, $\beta$, $\gamma$ is the bit string in subsystem $a$, $b$, $c$, respectively. And $\rho _{\alpha 0}^{X,ab}$ ($\rho _{\alpha 1}^{Y,ab}$) denotes the $\alpha 0th$ ($\alpha 1th$) diagonal term of the density matrix  $\rho^{X,ab}$ ($\rho^{Y,ab}$) with implementing an additional $\mathsf{RY}(-\pi/2)$ ($\mathsf{RX}(\pi/2)$) gate on qubit $b$ compared to $\rho^{ab}$.
Let's define $\overline{\beta }\equiv \beta \gamma$, therefor,
\begin{align}
 \rho _{\alpha \overline{\beta }}^{abc}=\rho _{\alpha 0}^{ab}\rho _{\left| 0 \right\rangle ,\overline{\beta }}^{bc}+\rho _{\alpha 1}^{ab}\rho _{\left| 1 \right\rangle ,\overline{\beta }}^{bc}+\frac{1}{2}(\rho _{\alpha 0}^{X,ab}-\rho _{\alpha 1}^{X,ab})(\rho _{\left| + \right\rangle ,\overline{\beta }}^{bc}-\rho _{\left| - \right\rangle ,\overline{\beta }}^{bc})  +\frac{1}{2}(\rho _{\alpha 0}^{Y,ab}-\rho _{\alpha 1}^{Y,ab})(\rho _{\left| +i \right\rangle ,\overline{\beta }}^{bc}-\rho _{\left| -i \right\rangle ,\overline{\beta }}^{bc}) \equiv \rho _{\alpha }^{ab,{{O}_{b}}}\circ \rho _{{{\psi }_{b}},\overline{\beta }}^{bc} 
\end{align}
where $\circ$ is a specially defined operator symbol for the computation between two sets, i.e.
   $\rho _{\alpha }^{ab,{{O}_{b}}}=\left\{ \rho _{\alpha 0},\rho _{\alpha 1},\rho _{\alpha 0}^{X},\rho _{\alpha 1}^{X},\rho _{\alpha 0}^{Y},\rho _{\alpha 1}^{Y} \right\} $ and
 $\rho _{{{\psi }_{b}},\overline{\beta }}^{bc}=\left\{ {{\rho }_{\left| 0 \right\rangle ,\overline{\beta }}},{{\rho }_{\left| 1 \right\rangle ,\overline{\beta }}},{{\rho }_{\left| + \right\rangle ,\overline{\beta }}},{{\rho }_{\left| - \right\rangle ,\overline{\beta }}},{{\rho }_{\left| +i \right\rangle ,\overline{\beta }}},{{\rho }_{\left| -i \right\rangle ,\overline{\beta }}} \right\} $.

In the fidelity characterization, one of the lower bound terms is
\begin{align}
   \text{Tr}\left( \rho \mathsf{ODD} \right)\text{=Tr}\left( {{\rho }_{ODD}}\mathsf{ODD}(X\to Z) \right)  =\frac{1}{{{2}^{\left\lceil n/2\right\rceil}}  }\left( 1+\sum\limits_{i=1}^{{{2}^{\left\lceil n/2\right\rceil}}-1}{\text{Tr}\left( {{\rho }_{ODD}}\text{S}{{\text{Z}}_{i}} \right)} \right) =\frac{1}{{{2}^{\left\lceil n/2\right\rceil}}}\left( 1+\sum\limits_{j=1}^{2^n}{{{p}_{j}}}{{\rho }_{ODD,j}} \right) 
\end{align}
where $n$ is the qubit number of ${\rho }$, and ${\rho }_{ODD}$ means implementing additional $\mathsf{RY}(-\pi/2)$ gates on odd qubits before measurement, and $\mathsf{ODD}(X\to Z)$ means replacing all $X$ by $Z$. $\text{S}{{\text{Z}}_{i}}$ means different expanded terms of the Pauli products, and ${p}_{j}$ equals $\pm1$ depending on $\text{S}{{\text{Z}}_{i}}$ which can be numerically calculated. Similarly, the other lower bound term is $\text{Tr}\left( \rho \mathsf{EVEN} \right)\text{=}{\left( 1+\sum\nolimits_{j}{{{q}_{j}}{{\rho }_{EVEN,j}}} \right)}/{{{2}^{\left\lfloor n/2\right\rfloor}}}$.

\begin{figure}
    \centering
    \includegraphics[width=0.96\textwidth]{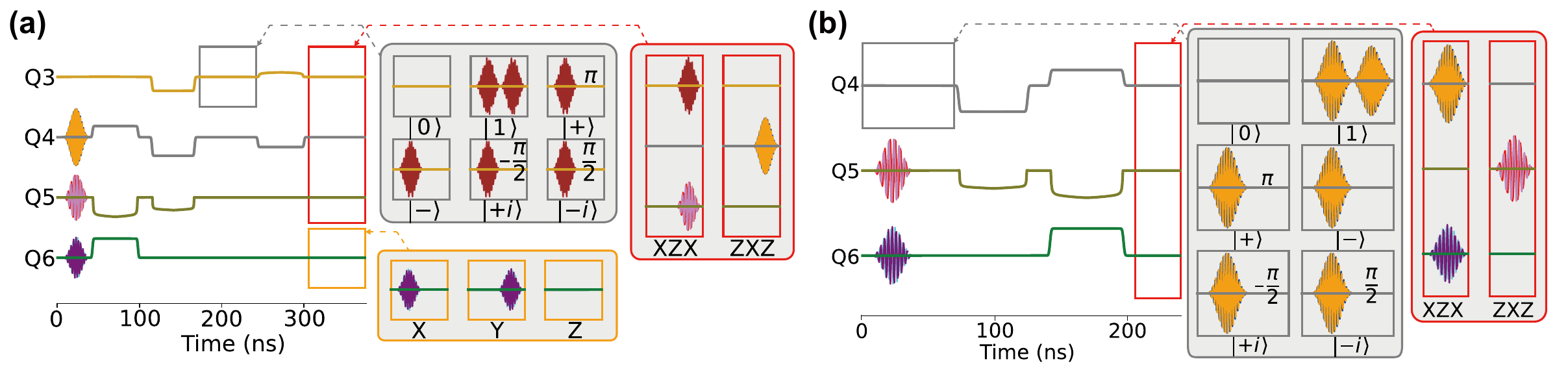}
    \caption{(a) 36 waveforms for 4-qubit subcircuits. The gray box shows the 6 initial state preparation waveforms (corresponding to $\ket{\psi_i}$ in the main text), in which three different virtual \textit{Z} are used. The red box shows the measurement bases waveforms of $XZX$ and $ZXZ$ for the first three qubits, and the orange box shows the measurement bases waveforms of $X$,$Y$,$Z$ for the fourth qubit. (b) 12 waveforms for 3-qubit subcircuits, with 6 initial state preparation waveforms in the gray box, and $XZX$ and $ZXZ$ measurements waveforms in the red box.}
    \label{fig:exp_pulse}
\end{figure}

In the case of 12 qubit simulation, the density matrix can be expressed as
\begin{align}
{{\rho }_{ODD}}=\sum\limits_{i}{{{c}_{i}}\text{T}{{\text{r}}_{{{O}_{i}}}}\left( \rho _{\left| + \right\rangle }^{XZX{{O}_{i}}} \right)\otimes \sum\limits_{j}{{{c}_{j}}\text{T}{{\text{r}}_{{{O}_{j}}}}\left( \rho _{{{\psi }_{i}}}^{ZXZ{{O}_{j}}} \right)\otimes \sum\limits_{k}{{{c}_{k}}\text{T}{{\text{r}}_{{{O}_{k}}}}\left( \rho _{{{\psi }_{j}}}^{XZX{{O}_{k}}} \right)\otimes }}}\rho _{{{\psi }_{k}}}^{ZXZ}
\end{align}
where $\rho _{\psi }^{XZX{{O}_{i}}}$ is the cutting circuit which is measured in basis $XZX{{O}_{i}}$ with the first qubit initiated in state $\psi$. The corresponding ${c}_{i}$, ${O}_{i}$ and ${\psi }_{i}$ are listed in the main text. Finally, the diagonal terms can be calculated by
\begin{align}
  {{\rho }_{ODD,\alpha \beta \gamma \eta }}=&\rho _{\left| + \right\rangle ,\alpha }^{XZX,{{O}_{i}}}\circ \left( \rho _{{{\psi }_{i}},\beta }^{ZXZ,{{O}_{j}}}\circ \left( \rho _{{{\psi }_{j}},\gamma }^{XZX,{{O}_{k}}}\circ \rho _{{{\psi }_{k}},\eta }^{ZXZ} \right) \right) \\ 
  {{\rho }_{EVEN,\alpha \beta \gamma \eta }}=&\rho _{\left| + \right\rangle ,\alpha }^{ZXZ,{{O}_{i}}}\circ \left( \rho _{{{\psi }_{i}},\beta }^{XZX,{{O}_{j}}}\circ \left( \rho _{{{\psi }_{j}},\gamma }^{ZXZ,{{O}_{k}}}\circ \rho _{{{\psi }_{k}},\eta }^{XZX} \right) \right)  
\end{align}
where $\left\{ \rho _{\psi ,\alpha }^{XZX,O},\rho _{\psi ,\alpha }^{ZXZ,O},\rho _{\psi ,\alpha }^{XZX},\rho _{\psi ,\alpha }^{ZXZ}|\psi \in \left\{ \left| 0 \right\rangle ,\left| 1 \right\rangle ,\left| + \right\rangle ,\left| - \right\rangle ,\left| +i \right\rangle ,\left| -i \right\rangle  \right\},O\in \left\{ X,Y,Z \right\} \right\}$ is the 48-circuit data in our experiments, whose experimental circuits are illustrated in Fig.~\ref{fig:exp_pulse} in the style of waveform.  3-qubit and 4-qubit linear-cluster states incidentally can be presented by $\left\{\rho _{\left| + \right\rangle ,\alpha }^{XZX,Z}, \rho _{\left| + \right\rangle ,\alpha }^{ZXZ,X}, \rho _{\left| + \right\rangle ,\alpha }^{XZX}, \rho _{\left| + \right\rangle ,\alpha }^{ZXZ}\right\}$.

\section{Mitigation Of Readout Noise}
Due to the imperfect readout, it is necessary to mitigate the readout noise for exact characterization of the state fidelity. A helpful and widely used method is transition matrix error mitigation (TMEM) method, which is implemented in our experiment. Let's denote the ideal vector of probabilities of a single qubit in an ideal device as ${{\left( p_{0}^{\text{ideal}},p_{1}^{\text{ideal}} \right)}^{\text{T}}}$ . The vector of probabilities ${{\left( p_{0}^{\text{exp}},p_{1}^{\text{exp}} \right)}^{\text{T}}}$  obtained in the experiment of a noisy device is given by
\begin{align}
\begin{pmatrix}
   p_{0}^{\mathrm{exp} }  \\
   p_{1}^{\mathrm{exp} }  \\
\end{pmatrix} =\left( \begin{matrix}
   {{f}_{00}} & 1-{{f}_{11}}  \\
   1-{{f}_{00}} & {{f}_{11}}  \\
\end{matrix} \right)\left( \begin{matrix}
   p_{0}^{\text{ideal}}  \\
   p_{1}^{\text{ideal}}  \\
\end{matrix} \right)
\end{align}
where ${f}_{00}$ and ${f}_{11}$ can be benchmarked experimentally as shown in Table \ref{tab:chip-performance}. So one can simply multiply ${{\left( p_{0}^{\text{exp}},p_{1}^{\text{exp}} \right)}^{\text{T}}}$ by the inverse of the matrix to reconstruct the ideal statistics. In terms of multi qubits, we benchmark the ${f}_{00}$ and ${f}_{11}$ of all qubits individually and obtain the transition matrices of each qubit. Then we use the direct product of transition matrices to correct our experimental probabilities.
However, the corrected probabilities vector of estimated statistics may contain non-physical negative elements. Therefor, we use the maximum likelihood method to find a proper density operator $\rho$ by maximizing the log-likelihood function $\sum\nolimits_{i}{{{p}_{i}}\log(\Tr(M_{i}\rho))}$, subject to $\Tr(\rho)=1$ and $\rho\ge 0$, where $M_{i}$ is the projection measurement operator and $p_{i}$ is the corresponding probability. Our experiment results are obtained depending on these error mitigation procedures.

\section{Complementary Data For 12-qubit Linear-Cluster State}

In the main text, we present experimental distributions from the 3-qubit and 4-qubit linear-cluster states in the $XZ$ measurement basis, and the expectations of terms from $\mathsf{ODD}$. 
Here, we give the more detailed data with the complementary data from the $ZX$ measurement basis and from the $\mathsf{EVEN}$ operator.
\red{The average distributions of two measurement bases are shown in Table~\ref{tab:3 qubit distribution} and Table~\ref{tab:4 qubit distribution} (the right-most bit is identified as the first bit).}



\begin{table*}[htbp]
\begin{tabular}{lcccccccccccccccc}
\toprule
  & 0000 & 0001 & 0010 & 0011 & 0100 & 0101 & 0110 & 0111 & 1000 & 1001 & 1010 & 1011 & 1100 & 1101 & 1110 & 1111 \\
\midrule
$P^{XZ}_{the}$ & 0.25 & 0 & 0 & 0 & 0 & 0 & 0 & 0.25 & 0 & 0 & 0 & 0.25 & 0.25 & 0 & 0 & 0 \\
$P^{XZ}_{exp}$ & 0.245 & 0 & 0 & 0 & 0 & 0 &  0 & 0.274 & 0.002 & 0 & 0 & 0.213 & 0.266 & 0 & 0 & 0.000 \\
$P^{ZX}_{the}$ & 0.25 & 0 & 0 & 0.25 & 0 & 0 & 0 & 0 & 0 & 0 & 0 & 0 & 0 & 0.25 & 0.25 & 0 \\
$P^{ZX}_{exp}$ & 0.226 & 0.000 & 0.000 & 0.248 & 0 & 0.004 & 0.009 & 0 & 0.024 & 0 & 0 & 0.003 & 0.005 & 0.245 & 0.235 & 0 \\
\bottomrule
\end{tabular}
\caption{Theory and experimental distributions of 4-qubit linear-cluster state in the $XZ$ measurement basis and $ZX$ measurement basis}
\label{tab:3 qubit distribution}
\end{table*}

\begin{table*}[htbp]
\begin{tabular}{lcccccccc}
\toprule
  & 000 & 001 & 010 & 011 & 100 & 101 & 110 & 111 \\
\midrule
$P^{XZ}_{the}$ & 0.5 & 0 & 0 & 0 & 0 & 0 & 0 & 0.5 \\
$P^{XZ}_{exp}$ & 0.444 & 0.010 & 0 & 0.026 & 0.043 & 0 & 0.010 & 0.468 \\
$P^{ZX}_{the}$ & 0.25 & 0 & 0 & 0.25 & 0 & 0.25 & 0.25 & 0 \\
$P^{ZX}_{exp}$ & 0.247 & 0 & 0 & 0.291 & 0.003 & 0.217 & 0.241 & 0 \\
\bottomrule
\end{tabular}
\caption{Theory and experimental distributions of 3-qubit linear-cluster state in the $XZ$ measurement basis and $ZX$ measurement basis}
\label{tab:4 qubit distribution}
\end{table*}

\begin{figure}
    \centering
    \includegraphics[width=0.96\textwidth]{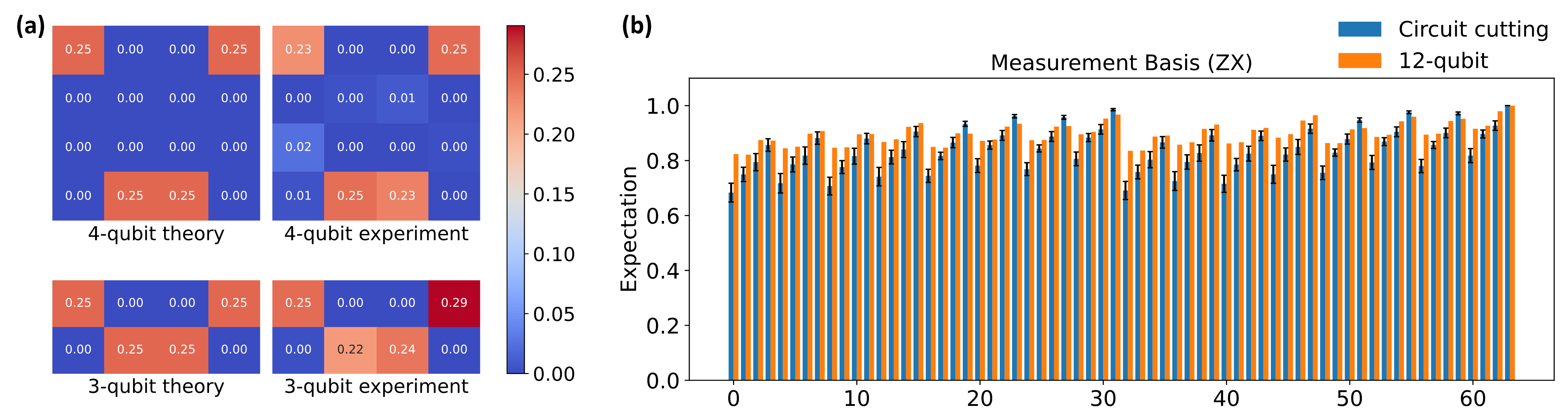}
    \caption{\textbf{(a)} The output distributions from $ZX$ measurement of the 4-qubit and 3-qubit linear cluster states.
    \textbf{(b)} Comparison of the expectations with $XZ$ measurement obtained from the 12-qubit circuit~\cite{gong_genuine_2019} (orange) and from the circuit-cutting scheme by running smaller quantum circuits (blue).
    Each bar corresponds to one specific observable in $\mathsf{EVEN}$ and there are 64 expectations for each group. The ideal values of all these expectations are one. }
    \label{fig:ZX_data}
\end{figure}

Fig.~\ref{fig:ZX_data}~(a) visualizes the probability distributions from the 4-qubit and 3-qubit linear-cluster state, and the theoretical distributions are also provided for a comparison.
\red{As in Fig.~3~(b) of the main text, each cell represents one bit string, and the number in the cell gives the output probability of obtaining that bit string.
In the 4-qubit case, the bit strings for the first row are 0000, 0001, 0010, 0011, and for the second row are 0100, 0101, 0110, 0111, and so on.
But note that, the bit strings are the measurement outcomes after the basis transformation.
That means, for example, in the $ZX$ measurement, 0011 will correspond to the output state $\ket{1}_1 \otimes \ket{-}_2 \otimes \ket{0}_3 \otimes \ket{+}_4$ (recall that in the bit-string representation, the right-most bit is identified as the first bit).}

\red{
Next, we explain how the bars in the Fig.~3~(c) in the main text are obtained. 
We only explain the 4-qubit case, but the 3-qubit case follows from the same procedure.
Recall the formula for the fidelity lower bound,
\begin{align}\label{eq:fidelity_lower_bound_supp}
    \Tr(\rho \dyad{\mathsf{LC}_n}) \geq \Tr(\rho \mathsf{ODD}_n) + \Tr(\rho \mathsf{EVEN}_n) - 1 \ .
\end{align}
For the 4-qubit LC state, its stabilizer group are spanned by $\{ X_1 Z_2, Z_1 X_2 Z_3, Z_2 X_3 Z_4, Z_3 X_4\}$, and its $\mathsf{ODD}$ and $\mathsf{EVEN}$ operators are given by,
\begin{align}
    \mathsf{ODD}_4 &= \left( \frac{I + X_1 Z_2}{2} \right) \left( \frac{I + Z_2 X_3 Z_4}{2} \right) = \frac{I + X_1 Z_2 + Z_2 X_3 Z_4 + X_1 X_3 Z_4}{4} \\
    \mathsf{EVEN}_4 &= \left( \frac{I + Z_1 X_2 Z_3}{2} \right) \left( \frac{I + Z_3 X_4}{2} \right) = \frac{I + Z_1 X_2 Z_3 + Z_3 X_4 + Z_1 X_2 X_4}{4} \ .
\end{align}
The nontrivial terms of $\mathsf{ODD}_4$ are shown in the first 3 bars in the left of Fig.~3~(c) in the main text, and those of $\mathsf{EVEN}_4$ are shown in the last 3 bars. 
The expectation value of $I$ is just a normalization condition, which will be always satisfied by a probability distribution.
The expectation values of the non trivial terms (the height of the bars) can be obtained from the distributions shown in Table~\ref{tab:4 qubit distribution}. 
For example, $\ev{X_1 Z_2}$ can be obtained from $P_{exp}^{XZ}$, and the observable is $1$ if the parity of the first two bits are even, and $-1$ if the parity of the first two bits are odd.
Therefore, we have,
\begin{align}
    \ev{X_1 Z_2} = 0.245 + 0.274 + 0.002 + 0.213 + 0.266 = 1 \ .
\end{align}
The value of other bars (observables) can be calculated with similar procedure.
After calculating the expectations of all the observables in $\mathsf{ODD}_4$ and $\mathsf{EVEN}_4$, the fidelity lower bound can be easily obtained from \eqref{eq:fidelity_lower_bound_supp}.
}

Fig.~\ref{fig:ZX_data}~(b) presents the expectations from the expansion terms in $\mathsf{EVEN}$ for the 12-qubit linear-cluster state, which can be obtained from the $ZX$ measurement.
\red{Here, we explain the labelling of the bars in Fig.~\ref{fig:ZX_data}~(b), and the same rule applies to Fig.~4~(a) in the main text.
First, as in the previous discussion, we expand \begin{align}
    \mathsf{EVEN}_{12} = \left( \frac{I + Z_1 X_2 Z_3}{2} \right) \left( \frac{I + Z_3 X_4 Z_5}{2} \right) \cdots \left( \frac{I + Z_{11} X_{12}}{2} \right)
\end{align} 
into $2^6 = 64$ terms. Each term can be represented by a binary vector (called the mask vector), with the 1's indicating the qubits that are acted nontrivially on.
For example, in the 4-qubit case, $Z_1 X_2 X_4$ is represented by $(1, 1, 0, 1)$ and $I$ is represented by $(0, 0, 0, 0)$.
Now, go back to the 12-qubit case and label every term in the expansion of $\mathsf{EVEN}_{12}$ by a binary vector of length 12. 
Each bar in Fig.~\ref{fig:ZX_data}~(b) is associated with one term, and hence one binary vector. 
From right to left, the binary vector is in a lexicographic order, which means that the 64-th bar is for $(0, 0, \cdots, 0)$ (or $I$), the 63-th bar is for $(0, 0, 0, 0, 0, 0, 0, 0, 0, 0, 1, 1)$ (or $Z_{11} X_{12}$), and so on.
We wrote a computer program to automate such labelling and its ordering~\cite{data_and_code}.
}

\red{The expectation values, or the heights of the bars, are obtained in the following way.
For the 12-qubit experiment, the expectation values are obtained from the measured output distributions in the $XZ$ and $ZX$ bases, just as what we discussed for the 3-qubit and 4-qubit case. 
As for the circuit-cutting experiment, the expectation values are obtained using Eq.~(8) in the main text.}
The average of these expectations in Fig.~\ref{fig:ZX_data}~(b), which is $\Tr(\rho \mathsf{EVEN})$, is 0.839 for the circuit-cutting experiment and 0.897 for the 12-qubit experiment.
Moreover, the average of the expectations from $XZ$ measurement (corresponding to Fig.~4~(a) in the main text), which is $\Tr(\rho \mathsf{ODD})$, is 0.895 for the circuit-cutting experiment and 0.717 for the 12-qubit experiment.
Combining the expectations for $\mathsf{ODD}$ and $\mathsf{EVEN}$, one can derive the fidelity lower bounds in the main text using \eqref{eq:fidelity_lower_bound_supp}.

As for the tradeoff, the circuit-cutting scheme saves qubits and allows for better control of the quantum system, at the cost of increasing the both the quantum and classical running time.
Both running times depend on the number of the cutting points, and the quantum running time is more expensive.
In the 12-qubit experiment, 25000 shots are used for each basis, and thus a total of 50000 shots are consumed; there is no repetition in the 12-qubit experiment.
In contrast, in the circuit-cutting experiment, there are 48 subcircuits to run, and 40000 shots are used for each subcircuits (1920000 shots in total); 
we perform 25 repeated experiments for the circuit-cutting scheme.
Although in a single experiment (repetition), the circuit-cutting scheme can consume far more shots than the direct implementation, it is nevertheless a valuable technique for pushing the limits of near-term quantum devices.

\bibliography{ref}